\documentclass{article}

\usepackage[main, final]{neurips_2025}

\usepackage[utf8]{inputenc} 
\usepackage[T1]{fontenc}    
\usepackage{hyperref}       
\usepackage{url}            
\usepackage{booktabs}       
\usepackage{amsfonts}       
\usepackage{nicefrac}       
\usepackage{microtype}      
\usepackage{xcolor}         
\usepackage{graphicx}
\usepackage{amsmath}
\usepackage{caption}

\title{Evaluating Operators for Acoustic Wave Simulation Correction}

\author{%
  Pascal Tribel\\
  \texttt{pascal.tribel@ulb.be} \\
  Université Libre de Bruxelles\\
  Machine Learning Group\\
  \And
  Gianluca Bontempi\\
  \texttt{gianluca.bontempi@ulb.be} \\
  Université Libre de Bruxelles\\
  Machine Learning Group\\
}

\begin{document}

\maketitle
\begin{abstract}
Correcting numerical dispersion artifacts from Finite Difference solvers is a well-identified challenge in computational wave physics, but existing approaches evaluate only a restricted family of CNN-based architectures and have been applied exclusively to the elastic wave equation. We instantiate the Deep Finite Difference framework on two-dimensional anisotropic acoustic wave propagation, pairing a fourth-order Finite Difference proxy with a Pseudo-Spectral reference over 27,000 heterogeneous velocity fields. We benchmark twelve correction architectures, from linear regression to Fourier Neural Operators, under a unified 10-fold cross-validation protocol.
\end{abstract}

\section{Introduction}\label{sec:existing_approaches}
Prediction of acoustic wave propagation is an important challenge in 
computational physics, with applications in seismic imaging~\citep{kamali_physics_informed_2024,zhang_seismic_2024}, 
fault detection~\citep{gu_2d_2019}, non-destructive testing~\citep{li_acoustic_2025}, 
and acoustic scene reconstruction~\citep{bianco_machine_2019}.

Numerical simulations suffer from two systematic sources of mismatch with reality: the acoustic wave equation neglects phenomena such as attenuation and medium deformation~\citep{blanch_1995}, and numerical solvers accumulate discretization errors over time~\citep{LIAO2018385,10.1190/1.1440470}. Prior work on elastic wave correction has sought to address this second problem through a common paradigm: two different numerical solvers generate a suboptimal solution and a reference one, and a neural network is trained to bridge them. \citet{gadylshin_ndm_2022} use U-Nets to correct 2D coarse-grid Finite Difference (FD) seismograms toward finer-grid references; \citet{siahkoohi_transfer_2019} apply a CNN to close the gap between low-order and higher-order FD solutions; \citet{rincon_neural_2025} use Fourier Neural Operators (FNO)~\citep{li2021fno} to correct elastic wave fields from a low-resolution spectral solver to high-resolution targets. Beyond wave simulation, \citet{kossaczka_deep_2023} introduced Deep FDM for correcting FD outputs toward analytical solutions in the heat and Black-Scholes equations~\citep{bohner2009analytical}.

However, while ML-based correction of elastic wave simulations has attracted growing attention~\citep{gadylshin_ndm_2022,siahkoohi_transfer_2019,rincon_neural_2025}, to the best of the author's knowledge, this framework has not been applied to the acoustic wave equation. Furthermore, each prior study proposes and evaluates a single architecture in isolation, leaving open the question of which learning method is best suited for this correction task. No systematic benchmark exists to motivate the choice of one approach over another.

We address both gaps, focusing on the correction of numerical errors. We instantiate the \textit{solver-in-the-loop} framework~\citep{um2020sol} for enhancing the simulation of 2D anisotropic acoustic wave propagations, pairing a fourth-order FD proxy with a Pseudo-Spectral (PS) reference to generate 27,000 trace pairs, thus implementing the Deep Finite Difference Method of \citet{kossaczka_deep_2023} for the acoustic setting. The correction task is cast as operator learning: mapping spatiotemporal traces generated by one solver to those of another can be described as the learning of a mapping from one function space to another. We conduct the first systematic benchmark of correction architectures for this task, evaluating a range of methods under a unified 10-fold cross-validation protocol, from linear regression (LR) and K-Nearest Neighbors (KNN) to Convolutional Neural Networks (CNN) and FNOs. The error between FD and PS-generated traces is spectral in nature, structurally aligned with FNO's global Fourier-domain convolutions and at odds with CNN architectures relying on finite-support spatial kernels~\citep{neuralop4}. Moreover, \citet{neuralop3} establish that Neural Operators such as FNOs are universal approximators of function-to-function mappings, providing theoretical grounding for their use alongside standard architectures. We therefore expect FNOs to outperform other architectures on this task.

\section{Data generation and pair of solvers}\label{sec:data_generation}
We use the 2D anisotropic acoustic wave equation~\citep{10.1190/1.1440470}:
\begin{equation}\label{eq:acoustic_wave}
    \frac{\partial^2 p}{\partial t^2} = c^2 \nabla^2 p + s
\end{equation}
where $p(x,y,t)$ is the pressure field, $c(x,y)$ the wave speed, and $s(x,y,t)$ an external source. We solve Equation~\eqref{eq:acoustic_wave} with two simulators on a 2D periodic uniform grid with $N_t = 256$ time steps, and we store the pressure evolution through time at the top surface, yielding trace pairs $(y_{\text{FD}}, y_{\text{PS}})$.

The reference uses Fourier Pseudo-Spectral discretization~\citep{Wise_2020,fornberg1996practical} with a second-order leapfrog time update:
\begin{equation}\label{eq:PSUpdate}
    p^{t+\delta t} = 2p^{t} - p^{t-\delta t}
        + c^{2}\delta t^{2}
          \Bigl(
            \mathcal{F}^{-1}\!\bigl[(ik_x)^{2}\mathcal{F}[p^{t}]\bigr] + \mathcal{F}^{-1}\!\bigl[(ik_y)^{2}\mathcal{F}[p^{t}]\bigr]
          \Bigr)
        + \delta t^{2}s^{t}
\end{equation}
The proxy uses a fourth-order centered FD stencil with $\mathcal{O}(\delta x^{4})$ spatial truncation error~\citep{fourth_order_fd,tam1993dispersion} and second-order leapfrog:
\begin{multline}\label{eq:FDUpdate}
    p^{t+\delta t} = 2p^{t} - p^{t-\delta t} + \delta t^{2}s^{t}
    + c^{2}\delta t^{2}
       \Biggl(
         \frac{-p_{x-2,y} + 16p_{x-1,y} - 30p_{x,y}
               + 16p_{x+1,y} - p_{x+2,y}}{12\delta x^{2}}\\
           + \frac{-p_{x,y-2} + 16p_{x,y-1} - 30p_{x,y}
               + 16p_{x,y+1} - p_{x,y+2}}{12\delta y^{2}}
       \Biggr)
\end{multline}

\begin{figure}[ht!]
    \centering
    \includegraphics[width=0.9\linewidth]{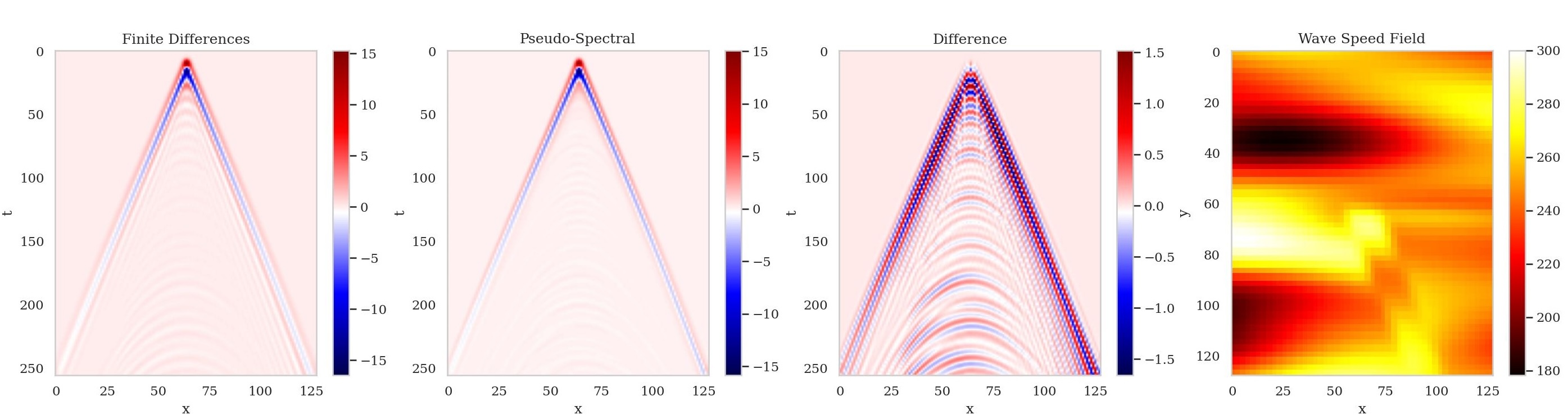}
    \caption{Sample example: (1) $y_{\mathrm{FD}}$, (2) $y_{\mathrm{PS}}$, (3) point-wise difference, and (4) velocity field (darker means slower).}
    \label{fig:sample_example}
\end{figure}

Heterogeneous velocity fields are extracted from \citet{NEGAHDARI2025134518}. The external forcing is a Ricker wavelet~\citep{10.1093/gji/ggu384} modulating a 2D Gaussian spatial envelope centered at the top surface midpoint. Both solvers use periodic boundary conditions: this is structurally required by the PS solver and eliminates absorbing-layer reflections so that the pairs of traces isolate FD dispersion error. We generate 27,000 pairs. Their difference is nearly symmetric (mean $5.2\times10^{-7}$, std $0.24$). The Root Normalized Mean Squared Error (RNMSE):
\begin{equation}\label{eq:rnmse}
    \text{RNMSE} = \sqrt{\frac{\frac{1}{N}\sum^N (\hat{y}-y_{\text{PS}})^2}{\sigma^2_{y_{\text{PS}}}}}
\end{equation}
between the traces in the whole dataset is $0.3856$, indicating substantial structured discrepancy. The error concentrates near the primary wavefront, consistent with fourth-order FD phase-velocity mismatch. A representative pair and corresponding velocity field are shown in Figure~\ref{fig:sample_example}.

\section{Correction Models}\label{sec:models}
All settings are evaluated using exhaustive 10-fold cross-validation~\citep{nti2021performance} on 27,000 samples with the RNMSE.
We compare Principal Components Regression (PCR)~\citep{pca,Massy01031965}\footnote{512 components in the PC space}, LR~\citep{su2012linear}, KNN~\citep{kramer2013k}\footnote{10 neighbors}, Extra Trees (ET)~\citep{geurts2006extremely}, and neural networks including MLP\footnote{One hidden layer of size 100, ReLU activation function}, CNN1d\footnote{Four hidden layers, kernel size of 3, ReLU activation function}, CNN2d\footnote{6 hidden layers, kernel size of 3, ReLU activation function}, UNet\footnote{6 hidden layers, skip connections by addition, ReLU activation function}, and Fourier Neural Operators\footnote{32 modes, 16 hidden channels}. An initial PCR correction yielding $\hat{y}_\text{PCR}$ is applied to the FD traces, correctly performing the PCA in the cross-validation pipeline to avoid data leakage. Each architecture then predicts $y_\text{PS}$ from $(\hat{y}_\text{PCR}, y_\text{FD})$. PCR, LR, KNN, and ET also perform the task directly on $y_\text{FD}$ as a baseline. PCR, LR, KNN, ET, and MLPs are implemented in scikit-learn~\citep{scikit-learn}; remaining architectures use PyTorch~\citep{NEURIPS2019_9015} and combine inputs as $\hat{y} = \alpha y_\text{FD} + \beta y_\text{PCR} + \gamma y_\text{NN}$, with $\alpha, \beta, \gamma$ optimized by gradient descent, and $y_\text{NN}$ being the output of the neural architecture. To assess the usefulness of the initial PCR guess in the correction task, we also use the three non-neural network methods to perform the correction task directly on $y_\text{FD}$.

\section{Results and Discussion}\label{sec:discussion}
\begin{figure}[ht!]
            \centering
            \includegraphics[width=0.7\linewidth]{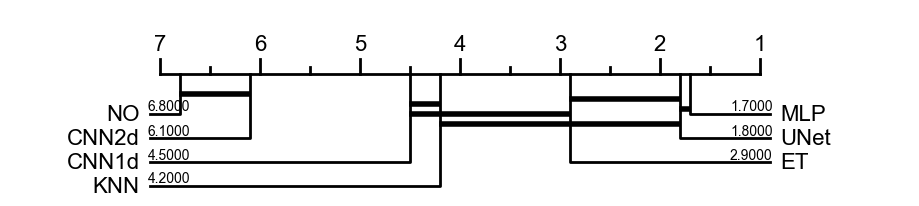}
            \caption{Critical Differences of the ML architectures. Cliques of non-significantly different results are obtained using the Wilcoxon Signed Rank test with Holm correction.}
            \label{fig:cddiagram}
    \end{figure}

\begin{figure}[ht!]
        \centering
        \includegraphics[width=0.8\linewidth]{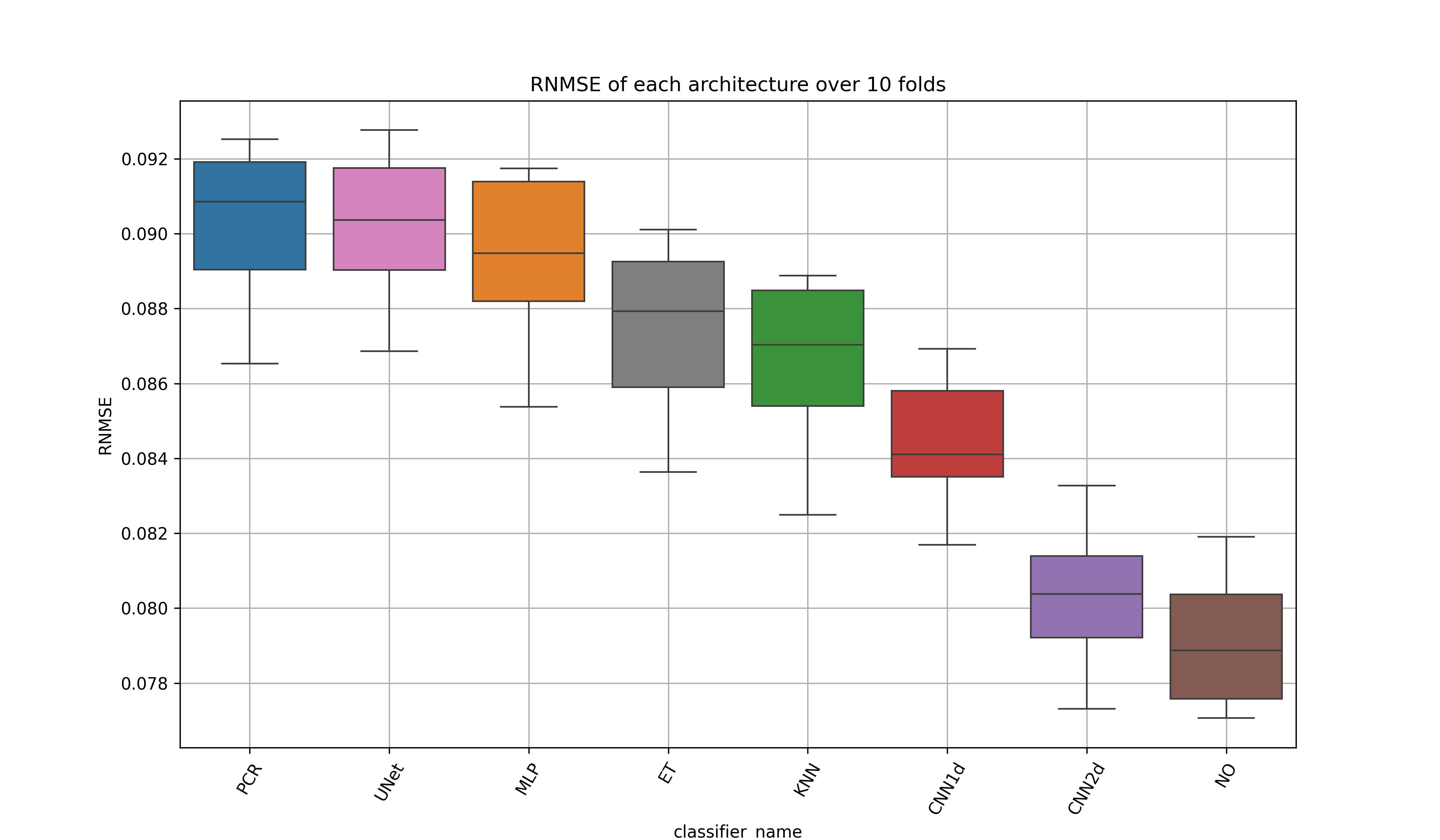}
        \caption{Distribution of the RNMSE for the selected architectures.}
        \label{fig:results_2}
    \end{figure}

    Table~\ref{tab:results} and Figure \ref{fig:results_2} report the 10-fold cross-validated RNMSE for each architecture. The last two columns show the p-value associated with the Wilcoxon test. Figure \ref{fig:cddiagram} shows the Critical Differences Diagram of the architectures that perform better than the initial PCR correction, where non-significant statistical differences are shown by bold horizontal lines and are obtained using Wilcoxon signed-ranked test with Holm correction~\citep{IsmailFawaz2018deep}. Architectures performing worse than the initial PCR pipeline (that is, the architectures correcting the trace directly without using the initial PCR correction) were discarded for readability.
    \begin{figure}[ht!]
        \centering
        \begin{tabular}{lrrrr}\hline
         & mean & std & p-value \\\hline
        Initial error & 0.3856 & & \\\hline
        PCR & 0.090359 & 0.001985 & 0.001953 \\
        UNet & 0.090176 & 0.001998 & 0.001953 \\
        MLP & 0.089523 & 0.002094 & 0.001953 \\
        ET & 0.087472 & 0.002169 & 0.001953 \\
        KNN & 0.086638 & 0.002117 & 0.001953 \\
        CNN1d & 0.084437 & 0.001643 & 0.001953 \\
        CNN2d & 0.080424 & 0.001803 & 0.013672 \\
        NO & 0.079006 & 0.001678 & \\
        \bottomrule
        \end{tabular}
        \captionof{table}{10-fold cross-validation performance, sorted by decreasing RNMSE. \textit{p-value} and \textit{r}: Wicoxon test's p-value against NO.}
        \label{tab:results}
    \end{figure}
    
    We discuss the following conclusions.
    \paragraph{All ML pipelines reduce the dispersion mismatch.}
    Every PCR-correcting pipeline reduces the mismatch below the baseline PCR error, confirming that a non-linear relationship exists between FD- and PS-generated traces that ML can exploit. Pipelines that omit the PCR initial guess perform substantially worse, supporting the hypothesis that removing the linear component first facilitates non-linear correction. Their results are not included in Table \ref{tab:results} nor in Figure \ref{fig:results_2} for the sake of readability.
    \paragraph{Statistical significance.} The results shown by the pairwise Wilcoxon test show a low \textit{p-value} for most architectures, but since the sample size is small ($n=10$), the minimal p-value computable is reached for most architectures.
    \paragraph{2D convolutions outperform 1D convolutions.}
    The dispersion error primarily manifests as a temporal phase shift along individual receiver traces~\citep{tam1993dispersion}. However, the spatial dependent nature of the acoustic wave explains the marginal advantage of \texttt{CNN2d} over \texttt{CNN1d}: a 2D temporal convolution captures both the spatial and temporal correction axes than a 1D kernel.
    \paragraph{FNOs achieve the best correction performance.}
    While the FNO achieves statistically significant improvement, the absolute margin over CNN2d is modest, especially since Figure \ref{fig:cddiagram} shows no significant statistical difference between the two architectures. Practitioners should weigh this against the FNO's greater variance and parameter count for their specific use case. This does not allow to check the validity of the theoretical argument of Section~\ref{sec:existing_approaches}, which hypothesized that the FNO's global Fourier-domain convolutions are structurally aligned with the spectral nature of the dispersion error, whereas CNN-based architectures are limited by the locality of their receptive fields~\citep{neuralop4}.

\section{Conclusion}\label{sec:conclusion}
    We introduced the first systematic benchmark for correcting numerical dispersion artifacts in 2D anisotropic \textit{acoustic} wave simulation, a setting previously unaddressed in the literature. Our contributions are twofold. We instantiated the Deep FDM on the acoustic wave equation, generating 27,000 anisotropic trace pairs from a fourth-order FD proxy and a PS reference. This constitutes the first publicly benchmarked dataset for acoustic solver correction. We evaluated twelve correction architectures, ranging from LR and KNN to U-Nets and FNOs, under a unified 10-fold cross-validation protocol. This is the first benchmark to include classical ML baselines alongside deep learning methods. All pipelines incorporating a PCR initial guess substantially outperform those operating directly on raw Finite Difference traces, confirming that removing the linear dispersion component first is critical. Among neural architectures, the Fourier Neural Operator achieves the best correction performance, although a larger number of experiments is required to show a significant difference. This is left for further work. While~\citet{rincon_neural_2025} similarly apply FNOs to wave solver correction, a direct comparison is precluded by fundamental differences in target equation (elastic vs. acoustic), solver pairing, and dataset, and we therefore treat their work as contextual motivation rather than a competing baseline. Future work could extend this framework to 3D simulations, investigate alternative NO architectures, and address the remaining gap between numerical solvers and physical reality. 

The experiment codes and architectures details are available for downloading \textit{\href{https://github.com/pascaltribel/does_knowledge_help}{on Github}}.

\subsubsection*{Acknowledgments} Pascal Tribel and Gianluca Bontempi are affiliated with \textit{TRusted AI Labs} (TRAIL). This publication benefits from the support of the Walloon Region to Pr. G. Bontempi as part of the funding for the FRFS‑WEL-T strategic axis. Gianluca Bontempi is supported by the Service Public de Wallonie Recherche under grant number 2010235-ARIAC by \textit{DigitalWallonia4.ai}, by the WEL Research Institute, Wavre, Belgique. Computational resources have been provided by the \textit{Consortium des Equipements de Calcul Intensif} (CECI), funded by the \textit{Fonds de la Recherche Scientifique de Belgique} (F.R.S.-FNRS) under Grant No. 2.5020.11 and by the Walloon Region. The author has no competing interests to claim.

\bibliographystyle{splncs04nat}
\bibliography{bibliography-4}
\end{document}